\documentclass[11pt]{article}
\usepackage{moriond,epsfig}

\def\be{\begin{equation}}
\def\ee{\end{equation}}
\def\bea{\begin{eqnarray}}
\def\eea{\end{eqnarray}}

\begin{document}
\vspace*{4cm}
\title{First Oscillation Results for the T2K Experiment}

\author{ M. Hartz, for the T2K Collaboration }

\address{Department of Physics, University of Toronto, 60 St. George St.,\\
Toronto, M4V 2B8, Canada}

\maketitle\abstracts{T2K is a long baseline high intensity neutrino oscillation
 experiment employing an off-axis design to search for the as yet unobserved
 appearance of $\nu_e$ neutrinos in a $\nu_{\mu}$ beam.  The neutrino beam originates 
at the J-PARC facility in Tokai, Japan and the Super-Kamiokande (SK) detector, 
located 295 km away, measures the composition of the oscillated beam.  The SK data 
are searched for an excess of $\nu_e$, constraining the allowed parameter space of 
$sin^2(2 \theta_{13})$, the parameter governing the amplitude of oscillations from 
$\nu_{\mu}$ to $\nu_e$.  This amplitude is of particular interest since it also 
modulates the amplitude of CP violating terms in the lepton mixing matrix.  
This paper presents results from the first T2K physics run in 2010 with $3.23\times10^{19}$
 protons on target.}

\section{Introduction}
In the three flavor oscillation model, neutrino mixing is parameterized by three mixing 
angles, $\theta_{12}$, $\theta_{23}$ and $\theta_{13}$ and a CP violating phase $\delta_{CP}$.
Additionally the oscillation probabilities depend on the mass squared differences between
the neutrino mass eigenstates, $\Delta m^{2}_{12}$ and $\Delta m^{2}_{23}$.
The mixing through $\theta_{12}$ has been well constrained by solar~\cite{Aharmim:2005gt} and
reactor~\cite{Abe:2008ee} experiments, while mixing through $\theta_{23}$ has been constrained by 
atmospheric~\cite{Ashie:2005ik} and accelerator based~\cite{Ahn:2006zza}~\cite{Michael:2006rx} experiments.
Searches for oscillations depending on $\theta_{13}$ have so far been inconclusive, but measurements by the 
CHOOZ~\cite{Apollonio:1999ae} and MINOS~\cite{Adamson:2009yc} experiments place upper limits on its value, 
$\sin^{2}(\theta_{13})<0.12-0.15$ at 90\% C.L..

The T2K (Tokai to Kamioka) experiment is a long baseline experiment designed with the primary
goal of searching for the appearance of electron neutrinos in a muon neutrino beam to 
measure the mixing angle $\theta_{13}$.  To leading order, the oscillation probability is:
\begin{equation}
P(\nu_{\mu} \rightarrow \nu_{e}) \approx sin^{2}(\theta_{23})sin^{2}(2\theta_{13})sin^{2}(\frac{\Delta m^{2}_{23}L}{4E_{\nu}})
\end{equation}
If the mixing angle $\theta_{13}$ is found to be non-zero, then the observation of CP violation in neutrino mixing
will be possible, and T2K will play an important role in searching for it.  This paper describes the first search for 
$\nu_{e}$ appearance at T2K.

\section{The T2K experiment}
The T2K experiment is described in detail elsewhere.~\cite{Abe:2011ks}  A brief description of the
experiment follows.
The T2K muon neutrino beam is produced when 30 GeV protons from the J-PARC accelerator facility collide 
with a 90 cm graphite target.  Positively charged particles (predominantly pions) produced in the collisions 
are focused by three magnetic horns and allowed to decay in a 96 m long decay volume.  The decay of the
$\pi^{+}$ hadrons produces a beam of $\nu_{\mu}$.  Decays of muons and kaons contaminate the 
beam with $\nu_{e}$ at the level of 1\%.  

T2K employs two near detectors located 280 m from the graphite target to measure the properties of the 
un-oscillated beam, and a far detector, the Super-Kamiokande (SK) detector, located 295 km away to measure
the oscillated beam.  SK sits 2.5$^{\circ}$ from the axis of the neutrino beam.
This off-axis angle takes advantage of the decay kinematics of pions
to produce a narrow band beam at the off-axis detector that peaks at the energy where neutrino oscillations
are expected.~\cite{Beavis:bnl52459}   

The INGRID near detector consists of 16 modules, 14 of which are arranged in a cross configuration, 
centered on the beam axis.  These modules consist of iron and scintillator layers and measure the neutrino 
rate and profile on the beam axis direction.  The ND280 off-axis near detector is located off the beam
axis in the same direction as SK and is used to measure the properties of the un-oscillated off-axis
beam.  ND280 consists of a number of sub-detectors, but for the analysis presented here, the Fine Grained 
Detectors (FGDs) and Time Projection Chambers (TPCs) are used.  The two FGDs consist of scintillator bars, 
with the second also including water targets.  Their 2.2 tons of mass provide the target material for neutrino 
interactions,
and the scintillation light is read out to reconstruct particle tracks near the interaction vertices.  The
three TPCs measure the momentum of charged particles in ND280's 0.2 T magnetic field to better than 10\% at
1 GeV/c.  They also provide dE/dx measurements with $<10\%$ resolution for particle identification.

The SK detector is a 50 kton water Cherenkov detector that consists of an inner (ID) and outer (OD) detector.
The OD is used to veto events that enter or exit the ID.  Neutrino interactions taking place in the
22.5 kton fiducial volume of the ID are detected by the Cherenkov light from charged interaction products
produced above threshold.  Photo-multiplier tubes instrumenting the walls of the ID image the 
Cherenkov light rings and the properties of the rings are used to reconstruct the particle type, energy and
vertex position.

The primary neutrino interaction mode that is of interest for T2K is the charged-current quasi-elastic (CCQE)
interaction, where a charged lepton and recoil nucleons are the only final state particles.  This interaction
mode is significant at T2K energies and allows for the approximate reconstruction of the neutrino 
energy if the charged lepton kinematics and neutrino beam direction are known.  An important background
interaction mode for T2K is the neutral-current $\pi^{0}$ (NC$\pi^{0}$) mode. Here the final state includes
the undetected neutrino, a $\pi^{0}$ and recoil nucleons. The $\pi^{0}$ decays to two photons that can be
misidentified as a single electron in the SK detector.  

The measurement described in this paper uses data accumulated with $3.23\times10^{19}$ protons on target from
January through June of 2010.  

\section{Electron neutrino appearance analysis}
To measure $\theta_{13}$, T2K searches for an excess of $\nu_{e}$ candidate events observed at SK that
can be interpreted as $\nu_{\mu} \rightarrow \nu_{e}$ oscillations.  There
are two major sources of background $\nu_{e}$ candidates at SK that must be accounted for: intrinsic $\nu_{e}$
contamination of the beam from muon and kaon decays and non $\nu_{e}$ interactions that are reconstructed as
$\nu_{e}$, in large part consisting of NC$\pi^0$ interactions.  The background and oscillation signal predictions are
produced using model and data based simulations of the neutrino flux and interactions, as well as the 
constraint from an inclusive $\nu_{\mu}$ measurement made using the ND280 detector.  The data over simulation 
rate measured at ND280 is used to renormalise the SK prediction:
\begin{equation}
N^{exp}_{SK} = N^{data}_{ND280}/N^{MC}_{ND280} \times N^{MC}_{SK}
\end{equation}
By doing this, the neutrino rate prediction is constrained by the near detector data, and significant cancellations
in the neutrino flux uncertainties are realized. 
The SK $\nu_{e}$ selection
is applied to the simulation as well as the data, and the measured number of events compared to the prediction
provides a constraint on $\theta_{13}$.

\subsection{SK $\nu_e$ selection}
The selection criteria for $\nu_{e}$ candidates at SK was finalized before looking at the data to avoid 
bias.  Cuts were optimized for the relatively small expected sample size of T2K's initial data sets.
The selection looks for events with a single electron like ring that will be produced by the final state electron
in CCQE interaction of $\nu_{e}$.

The selection of SK $\nu_{e}$ candidates begins with the sample of neutrino interaction candidates 
that are fully contained in the ID with vertices in the fiducial volume.  A $>100$ MeV visible energy cut is
applied to reduce the backgrounds from neutral current interactions or electrons from muon decays.  The
candidates are required to have a single ring, and the ring must be identified as an electron like ring.  
Electron rings are identified by their ``fuzzy" edges compared to muon rings due to the electromagnetic 
scattering of the electron in the water.
No delayed activity can be observed in the detector as this is interpreted as electrons from muon decays.  
For each event, a $\pi^{0}$ mass is reconstructed under the two ring hypothesis, and if $m_{\pi^0}>105$ $MeV$/c$^{2}$
the event is rejected.  This cut removes background due to photons from $\pi^0$ decays.  Finally, 
the reconstructed energy of the $\nu_e$ candidate is required to be $<1250$ $MeV$ since the oscillation probability
peaks below $1000$ $MeV$.  This selection has an efficiency of 66\% for signal events with efficiency uncertainties
of 7.6\% and 15.8\% for signal and background respectively. 

\subsection{Flux prediction}
\begin{figure}
\vspace{0.4in}
\begin{center}
\psfig{figure=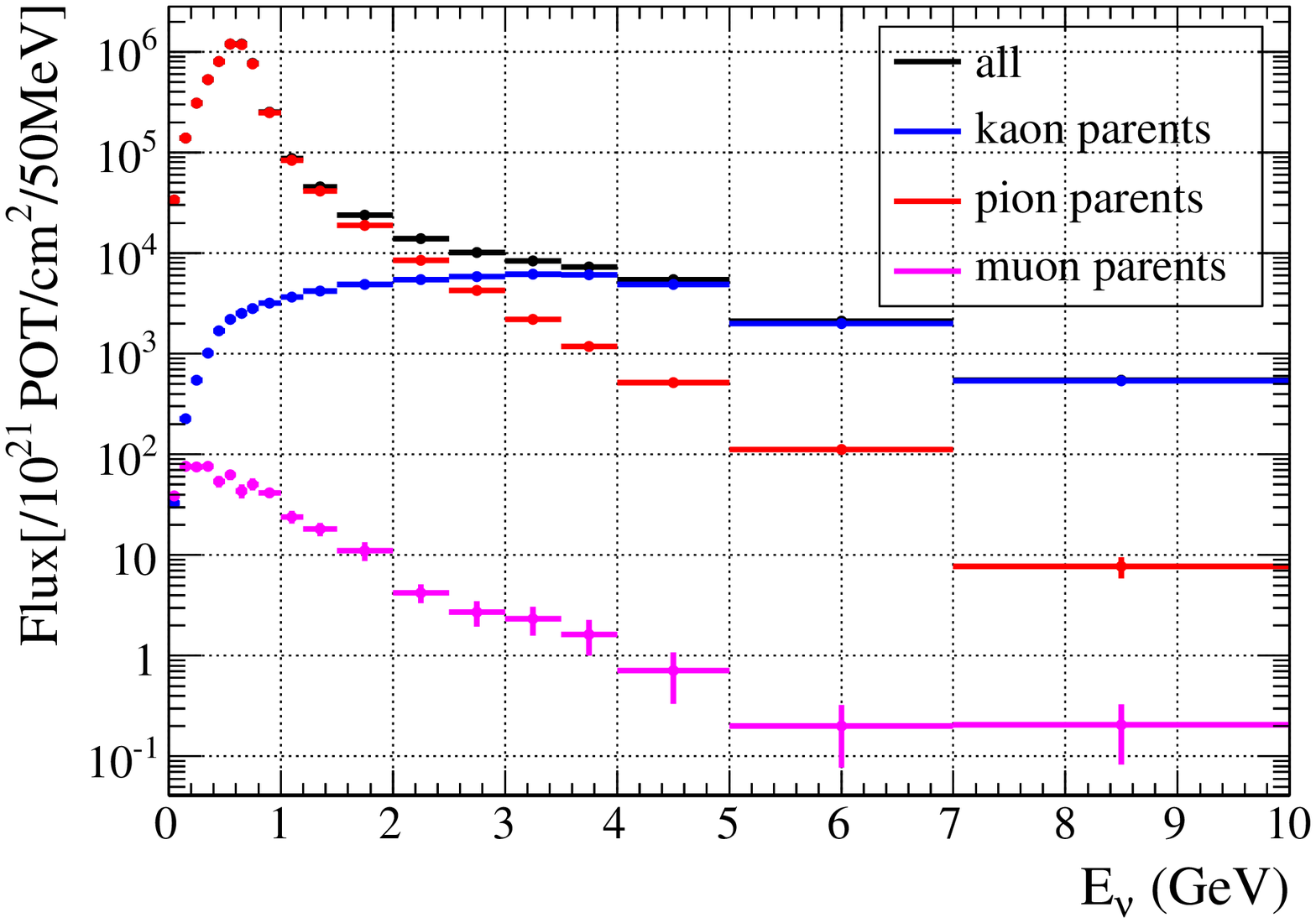,height=1.8in}
\hspace{0.4in}
\psfig{figure=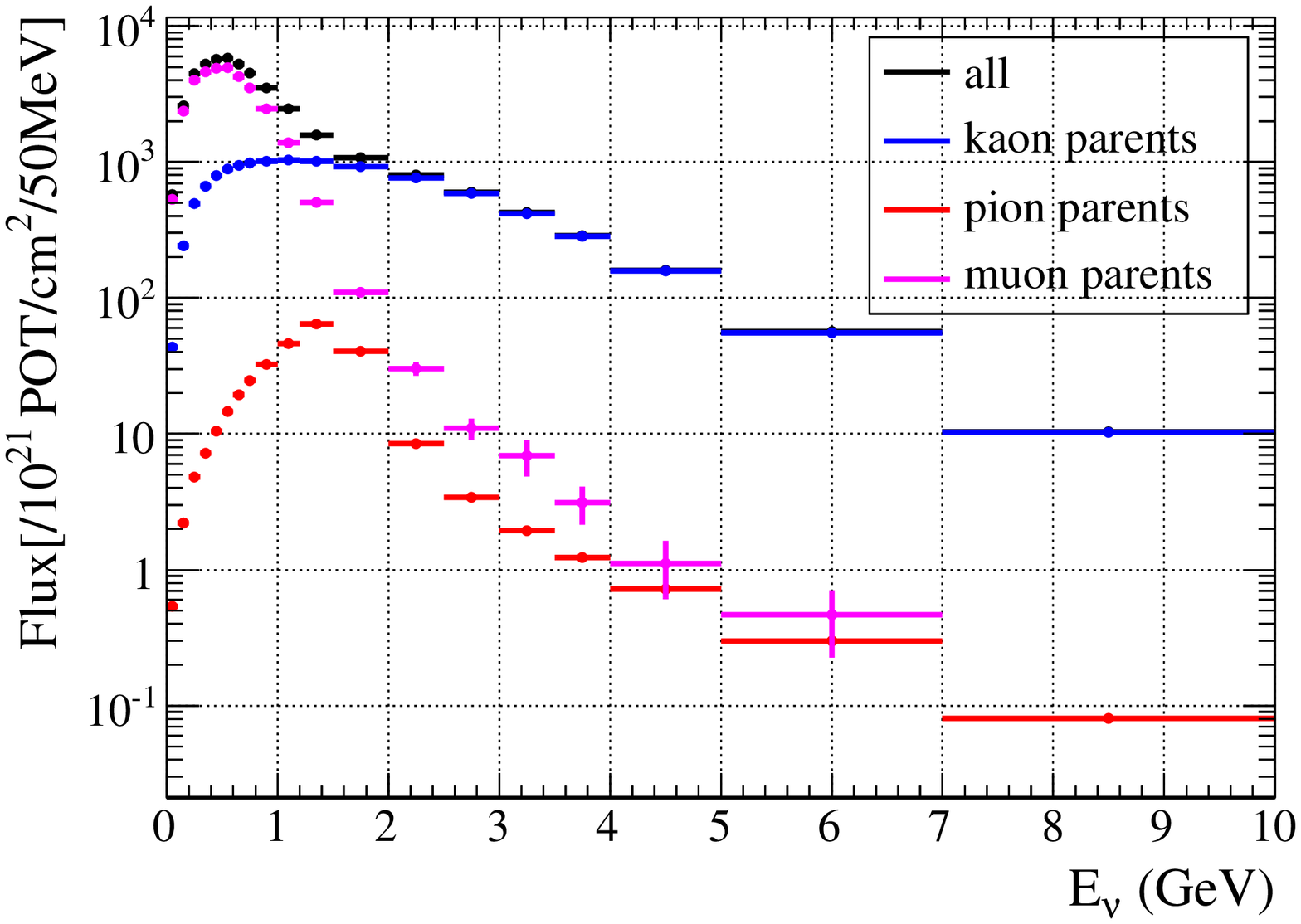,height=1.8in}
\end{center}
\caption{Predicted $\nu_{\mu}$ (left) and $\nu_{e}$ (right) fluxes at SK based on simulation. Error
bars represent the statistical uncertainty of the simulated flux.
\label{fig:flux}}
\end{figure}

The flux prediction is made from the simulation of protons interacting in the T2K target and the subsequent
propagation of secondary particles through the magnetic horns and decay volume until they decay to produce
neutrinos.  T2K proton beam monitor measurements are used to set the initial conditions for the
protons in the simulation.  The production of pions by proton interactions inside the target are modeled 
with data from the NA61 experiment~\cite{na61}, while other in-target interactions
are modeled with FLUKA.~\cite{Battistoni:2007zzb}~\cite{Ferrari:2005zk}  Propagation of particles outside the 
target is carried out with GEANT3~\cite{GEANT3} and hadron interactions are modeled with the GCALOR~\cite{GCALOR} 
package.  Fig.~\ref{fig:flux} shows the expected $\nu_{\mu}$ and $\nu_{e}$
neutrino fluxes seen by SK, broken down by the parent particle that produces the neutrino. The $\nu_{\mu}$ produced
in the $<1$ $GeV$ region of interest are predominantly from pion decays, while the $\nu_{e}$ contamination is
predominantly from muon decays.  The dominant sources
of uncertainty in the neutrino flux come from the production of pions and kaons in the interactions of protons, 
and the total flux uncertainty contributes a 9.2\% uncertainty to $\nu_{e}$ background candidate prediction.

\subsection{Neutrino interaction modeling}
The interactions of neutrinos are modeled with the NEUT~\cite{NEUT} neutrino interaction generator, while
the GENIE~\cite{GENIE} neutrino interaction generator is used for cross-checks.  The uncertainties on the 
neutrino interaction models are evaluated in three ways:
\begin{itemize}
\item Comparisons between models
\item Variations of parameters within models
\item Comparisons to data from the MiniBooNE~\cite{AguilarArevalo:2010zc} and SciBooNE~\cite{Kurimoto:2009wq}
~\cite{Nakajima:2010fp} experiments, as well as the SK atmospheric data set
\end{itemize}
The uncertainty on the SK $\nu_{e}$ candidate sample size from background sources due to neutrino interaction
uncertainties is 14.2\%.  The dominant sources of uncertainties are final state interactions of pions, and
the NC$\pi^0$ cross section. 

\subsection{ND280 inclusive $\nu_{\mu}$ measurement}
\begin{figure}
\begin{center}
\psfig{figure=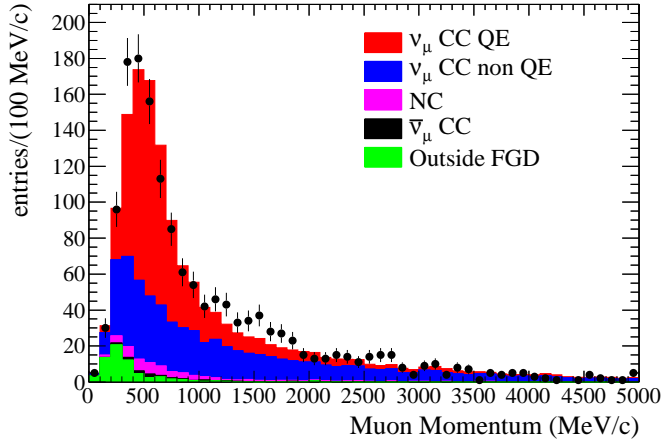,height=2.5in}
\end{center}
\caption{Muon momentum from the inclusive $\nu_{\mu}$ interaction data measured at ND280.  Error bars
represent the statistical uncertainty of the data points.
\label{fig:nd280mom}}
\end{figure}

The rate of $\nu_{\mu}$ charged current interactions is measured by ND280 using a sample of events 
where a negative track originates in one of the two FGDs and is tracked by the downstream TPC.  The TPC dE/dx
measurement is used to select muons and reject electrons, resulting in a sample that is $90\%$ $\nu_{\mu}$
charged current interactions, and $50\%$ CCQE.  Fig.~\ref{fig:nd280mom} shows the predicted distribution of 
reconstructed muon momentum compared to the measured distribution.  The ratio of data over the prediction for 
the full sample is:
\begin{equation}
N^{data}_{ND280}/N^{MC}_{ND280} = 1.061\pm0.028(stat.)^{+0.044}_{-0.038}(syst.)\pm0.039(phys. model)
\end{equation}
This ratio is used to renormalise the SK event rate predictions, and the uncertainties on this ratio 
are propagated into uncertainty on the predicted SK samples.

\subsection{SK $\nu_{e}$ prediction}
Using the flux prediction, neutrino interaction models and near detector measurement,
the background and signal expectations for $\nu_{e}$ candidates at SK are calculated.  Table~\ref{sk_pred} shows
the predictions for $3.23\times10^{19}$ p.o.t. and $\sin^{2}(2\theta_{13})=0.1$.  The background prediction
is $0.30\pm0.07(syst.)$ events.  The dominant sources of uncertainty come from the flux prediction, neutrino
interaction modeling and SK ring counting, particle ID and $\pi^0$ mass cuts.

\begin{table}
\begin{center}
\caption{SK $\nu_{e}$ candidate predictions for $3.23\times10^{19}$ p.o.t. and $\sin^{2}(2\theta_{13})=0.1$.}
\begin{tabular}{l|l|l}
\hline
Source & Events & Systematic Error \\ \hline 
Background                                   & 0.30 & 23.9\% \\
\hspace{0.1in}   Beam $\nu_{e}$ (85\% CCQE)  & 0.16 &    \\
\hspace{0.1in}   $\nu_{\mu}$ (95\% NC)       & 0.13 &   \\
\hspace{0.1in}  $\bar{\nu}_{\mu}$            & 0.01 &    \\
Signal $\nu_{e}$                             & 1.20 & 19.5\% \\ \hline
\end{tabular}
\end{center}
\label{sk_pred}
\end{table}

\subsection{SK data sample and interpretation}
\begin{figure}
\vspace{0.4in}
\begin{center}
\psfig{figure=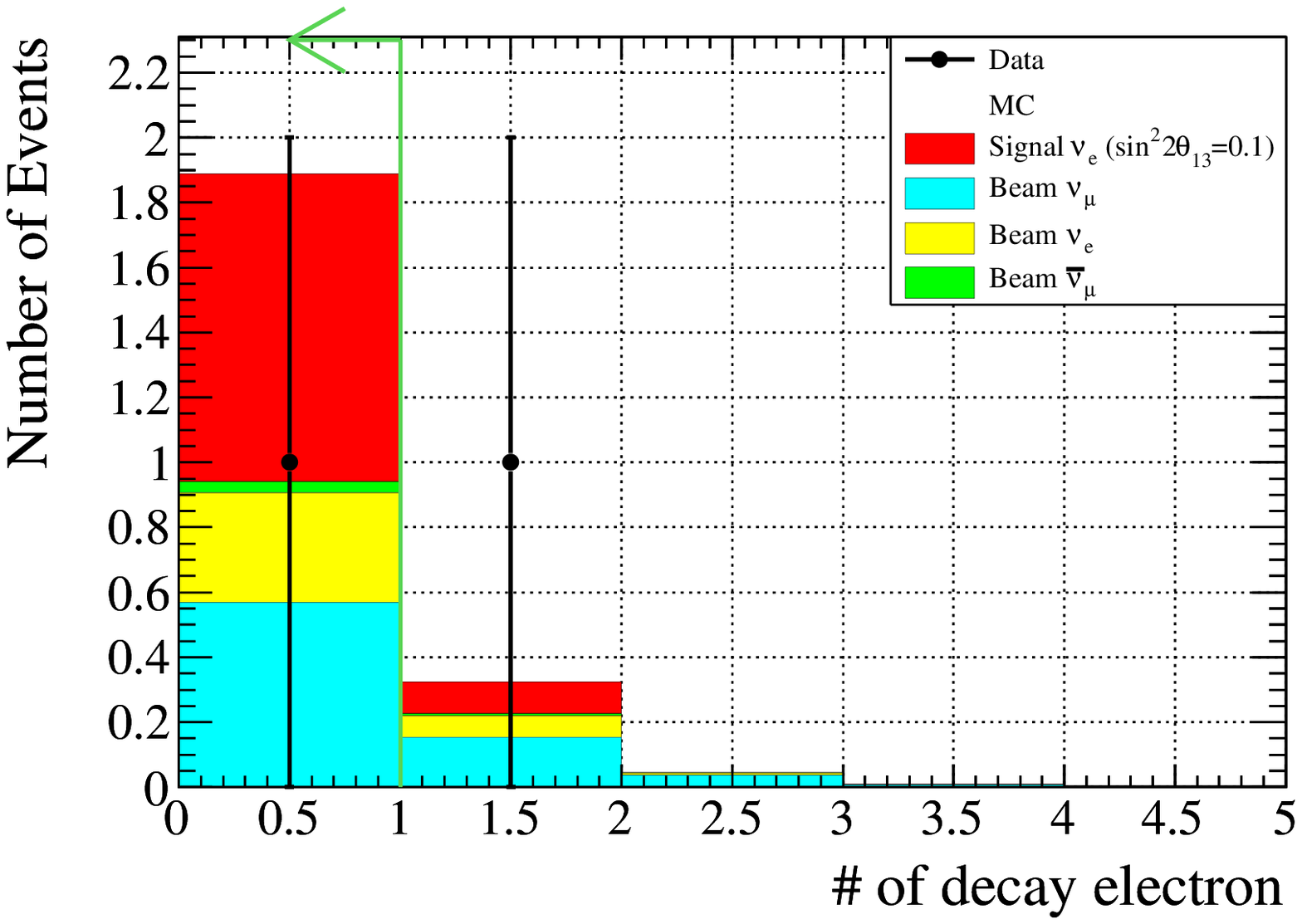,height=1.8in}
\hspace{0.4in}
\psfig{figure=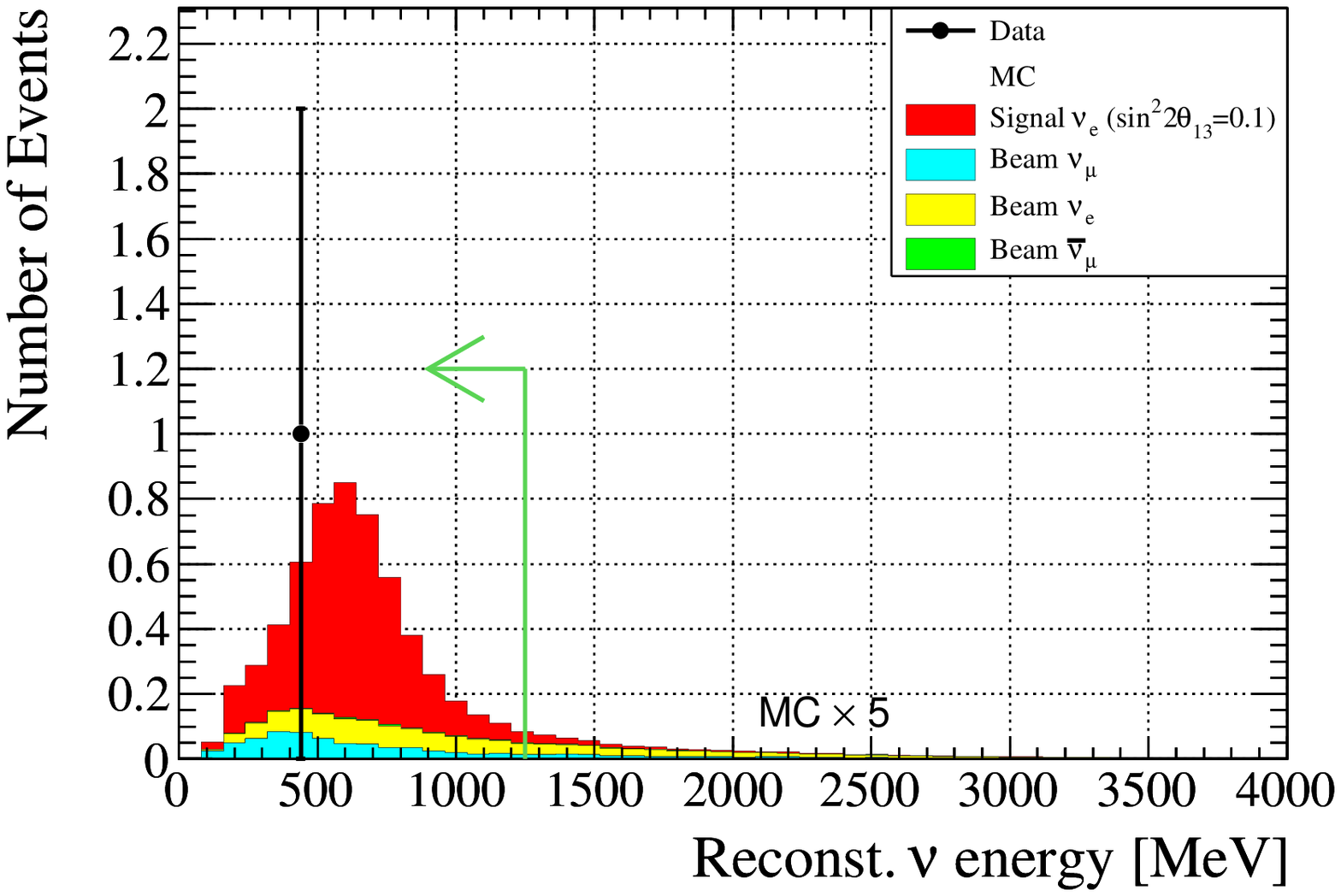,height=1.8in}
\end{center}
\caption{Data and predicted $\nu_{e}$ candidate samples at the decay electron (left) and 
reconstructed neutrino energy (right) cuts.
\label{fig:nue_cuts}}
\end{figure}

\begin{figure}
\begin{center}
\psfig{figure=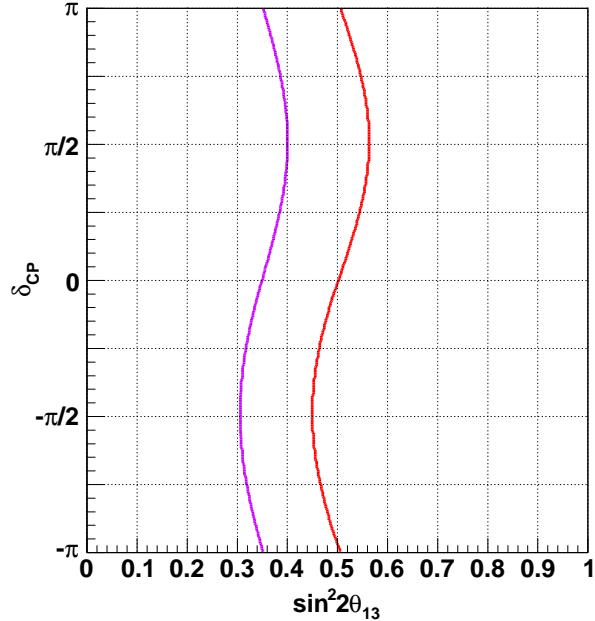,height=3.5in}
\end{center}
\caption{The 90\% C.L. upper limit (red) and sensitivity (magenta) for the T2K data set with
$\Delta m^{2}_{23}(>0) = 2.4\times10^{-3}$ $eV^2$ and $sin^{2}(2\theta_{23})=1.0$.
\label{fig:nue_limits}}
\end{figure}

The $\nu_{e}$ selection cuts are applied to the SK data and the resulting number of events is used
to place a constraint on $\theta_{13}$.  Fig.~\ref{fig:nue_cuts} shows the effect of the 
decay electron and reconstructed neutrino energy cuts on the data
and predicted distributions.  After all cuts are applied, one candidate event remains.  With this single event
and the background and signal predictions, 
limits on $sin^{2}(2\theta_{13})$ are calculated using the Feldman-Cousins~\cite{Feldman:1997qc} method.  The resulting
90\% C.L. limit for varying $\delta_{CP}$ are shown in Fig.~\ref{fig:nue_limits}.  For $\delta_{CP} = 0$,
$\Delta m^{2}_{23} = 2.4\times10^{-3}$ $eV^2$ and $sin^{2}(2\theta_{23})=1.0$ the 90\% C.L. upper limit
is found to be 0.5.  

\section{Conclusion}
T2K has carried out a search for $\nu_{e}$ appearance in a $\nu_{\mu}$ beam with data produced from
$3.23\times10^{19}$ protons on target.  In these data, T2K observes one $\nu_{e}$ candidate event at 
the SK detector when $0.30\pm0.07(syst.)$  events are expected from background sources.  T2K sets the 
upper limit $sin^{2}(2\theta_{13})<0.5$ at 90\% C.L. (for $\Delta m^{2}_{23} = 2.4\times10^{-3}$ $eV^2$ 
and $sin^{2}(2\theta_{23})=1.0$).  Although this first measurement from T2K does not yet challenge the
sensitivity of previous experiments' measurements, future T2K measurements will follow, with
four times the data set already available, promising interesting results in the near future.

%\begin{figure}
%\rule{5cm}{0.2mm}\hfill\rule{5cm}{0.2mm}
%\vskip 2.5cm
%\rule{5cm}{0.2mm}\hfill\rule{5cm}{0.2mm}
%\psfig{figure=filename.ps,height=1.5in}
%\caption{Radiative (off-shell, off-page and out-to-lunch) SUSY Higglets.
%\label{fig:radish}}
%\end{figure}

\end{document}